\documentclass[francais]{gretsi}

\usepackage[utf8]{inputenc}

\usepackage[T1]{fontenc}

\usepackage{amsmath, amssymb, amsfonts}

\usepackage[dvipsnames]{xcolor}
\usepackage{hyperref}
\hypersetup{
    colorlinks = true,
    urlcolor = black,
    linkcolor = Cerulean,
    citecolor = WildStrawberry,
}

\usepackage{subfig}

\usepackage{multicol}

\newcommand{\dt}{\operatorname{d}\!t}
\newcommand{\transpose}[1]{{#1}^\mathsf{T}}

\begin{document}

	\titre{Recursive KalmanNet : Analyse des capacités de généralisation d'un réseau de neurones récurrent guidé par un filtre de Kalman}

	\auteurs{ 
        \auteur{Cyril}{Falcon}{cyril.falcon@exail.com}{}
        \auteur{Hassan}{Mortada}{hassan.mortada@exail.com}{}
        \auteur{Mathéo}{Clavaud}{matheo.clavaud@exail.com}{}
        \auteur{Jean-Philippe}{Michel}{jean-philippe.michel@exail.com}{}
    }
	\affils{
        \affil{}{
            Exail -- Systèmes de Navigation et Applications\\ 
            34 rue de la Croix de Fer, 78100 Saint-Germain-en-Laye, France
        }
    }

	\resume{Le Recursive KalmanNet, récemment introduit par les auteurs, est un réseau de neurones récurrent guidé par un filtre de Kalman, capable d'estimer les variables d'état et la covariance des erreurs des systèmes dynamiques stochastiques à partir de mesures bruitées, sans connaissance préalable des caractéristiques des bruits. Cet article explore ses capacités de généralisation dans des scénarios hors distribution, où les dynamiques temporelles des mesures de test diffèrent de celles rencontrées à l'entraînement.}

	\abstract{The Recursive KalmanNet, recently introduced by the authors, is a recurrent neural network guided by a Kalman filter, capable of estimating the state variables and error covariance of stochastic dynamic systems from noisy measurements, without prior knowledge of the noise characteristics. This paper explores its generalization capabilities in out-of-distribution scenarios, where the temporal dynamics of the test measurements differ from those encountered during training.}

	\maketitle

    \section{Introduction}
    \label{sec:intro}

        L'estimation des états d'un système dynamique stochastique à partir de mesures bruitées est une thématique centrale en traitement du signal, avec de nombreuses applications variées, notamment en navigation inertielle. Le filtre de Kalman \cite{Kalman} s'est imposé comme une solution de référence \cite{Schmidt} grâce à sa structure analytique, facilitant son implémentation embarquée, et à son optimalité sous certaines conditions. Cependant, ces hypothèses sont rarement vérifiées en pratique, et le réglage du filtre, souvent conservateur en raison de la méconnaissance des bruits, en dégrade les performances. Des méthodes hybrides associant filtrage de Kalman et apprentissage profond ont été proposées \cite{KalmanNet, SplitKalmanNet, CholeskyKalmanNet} pour dépasser ces limitations.
        
        Nous avons récemment conçu le Recursive KalmanNet \cite{RecursiveKalmanNet}, un réseau de neurones récurrents guidé par un filtre de Kalman, destiné à estimer les variables d'état et modéliser la covariance des erreurs. Son évaluation montre qu'en présence de bruit de mesure non gaussien, il surpasse non seulement un filtre de Kalman configuré de manière réaliste, mais aussi les approches basées sur l'apprentissage profond.

        Cependant, bien qu'un réseau de neurones soit performant sur les données d'entraînement, il peut devenir médiocre sur des données inédites \cite{Generalization}. Cette étude explore les capacités de généralisation du Recursive KalmanNet lorsque le bruit de mesure varie entre l'entraînement et le test, un cas crucial en navigation inertielle. Nos résultats montrent que, bien que la précision reste élevée, les covariances estimées perdent en représentativité des erreurs empiriques.
        
        Après avoir présenté le filtre de Kalman et les méthodes profondes dérivées pour l'estimation des états (Section \ref{sec:estimation}), nous introduisons le Recursive KalmanNet (Section \ref{sec:recursive_kalmannet}) et évaluons ses capacités de généralisation (Section \ref{sec:generalisation}).

        \begin{figure*}[ht!]
            \centering

            \includegraphics[width=0.74\linewidth]{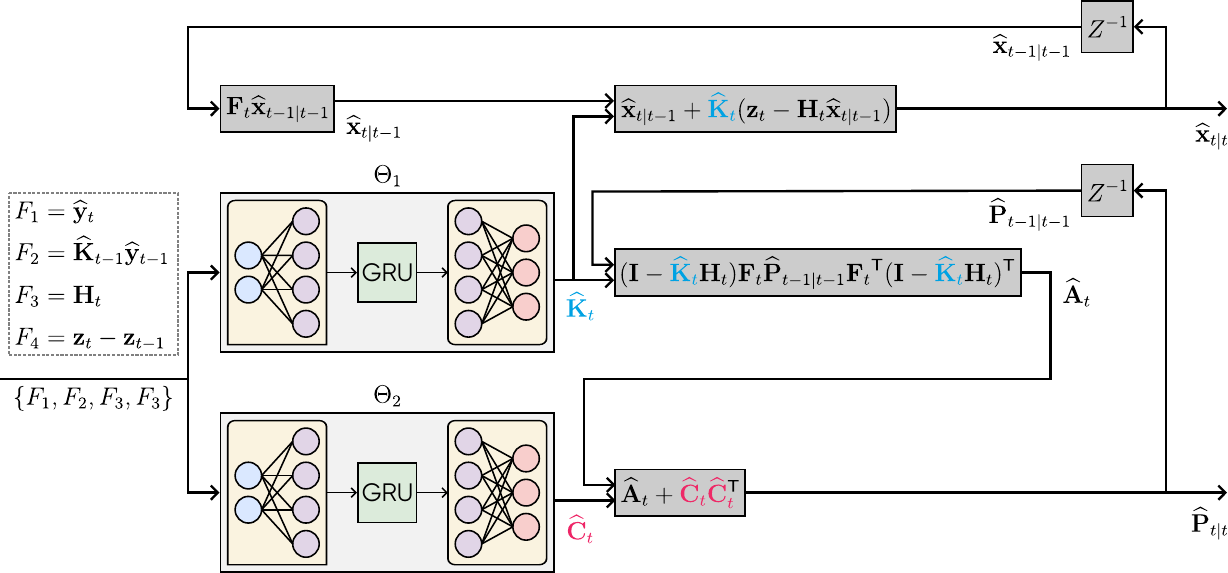}
            \caption{Architecture du Recursive KalmanNet, où $\Theta_1$ et $\Theta_2$ sont les paramètres des deux réseaux de neurones récurrents.}
            \label{fig:RKN_architecture}
        \end{figure*}

    \section{Estimation de variables d'état}
    \label{sec:estimation}

        Cet article porte sur l'estimation des variables d'état d'un système dynamique stochastique à partir de mesures bruitées. Nous nous restreignons aux systèmes dynamiques linéaires et discrets, dans lesquels les mesures sont des combinaisons linéaires des variables d'état.

        Dans la suite, nous désignons par $\mathbf{x}_t \in \mathbb{R}^m$ et $\mathbf{z}_t \in \mathbb{R}^n$ les vecteurs d'état et de mesure à l'instant $t$, et nous adoptons une représentation d'état définie par :
        \begin{subequations}
            \begin{align}
                \label{eq:propagation}
                \mathbf{x}_t & = \mathbf{F}_t \mathbf{x}_{t - 1} + \mathbf{v}_t, \\
                \label{eq:mesure}
                \mathbf{z}_t & = \mathbf{H}_t \mathbf{x}_t + \mathbf{w}_t.
        \end{align}
        \end{subequations}
        Avec $\mathbf{F}_t \in \mathcal{M}_{n,n}(\mathbb{R})$ et $\mathbf{H}_t \in \mathcal{M}_{n,m}(\mathbb{R})$, appelées matrices de propagation et de mesure, et $\mathbf{v}_t \in \mathbb{R}^m$ et $\mathbf{w}_t \in \mathbb{R}^n$, appelés bruits de propagation et de mesure, supposés indépendants, blancs, de moyennes nulles, et de covariances respectivement notées $\mathbf{Q}_t \in \mathcal{M}_{m,m}(\mathbb{R})$ et $\mathbf{R}_t \in \mathcal{M}_{n,n}(\mathbb{R})$.
        
        \subsection{Filtre de Kalman}
        \label{sec:kalman}

            Le filtre de Kalman (KF) \cite{Kalman} est un estimateur linéaire récursif qui fournit une solution analytique au problème d'estimation du vecteur d'état du système défini par les équations \eqref{eq:propagation}--\eqref{eq:mesure}. Il repose sur un schéma prédicteur-correcteur qui estime à la fois le vecteur d'état et la covariance de l'erreur d'estimation. Cet estimateur est optimal au sens de l'erreur quadratique moyenne lorsque les bruits de propagation et de mesure sont gaussiens, et reste le meilleur estimateur linéaire en présence de bruits non gaussiens.
            
           Une fois l'état initial $\widehat{\mathbf{x}}_{0 \vert 0}$ et la covariance de l'erreur $\mathbf{P}_{0 \vert 0}$ initialisés, les corrections du vecteur d'état et de la covariance de l'erreur au temps précédent sont propagées par \eqref{eq:propagation} pour prédire le vecteur d'état et la covariance de l'erreur au temps courant, selon :
            \begin{subequations}
                \begin{align}
                    \label{eq:prediction}
                    \widehat{\mathbf{x}}_{t \vert t-1} & = \mathbf{F}_t \widehat{\mathbf{x}}_{t-1 \vert t-1}, \\
                    \label{eq:covariance_prediction}
                    \mathbf{P}_{t \vert t-1}           & = \mathbf{F}_t \mathbf{P}_{t-1 \vert t-1} \transpose{\mathbf{F}_t} + \mathbf{Q}_t.
                \end{align}
            \end{subequations}
            Ensuite, la mesure au temps courant est utilisée pour corriger la prédiction du vecteur d'état et de la covariance de l'erreur. D'abord, l'innovation et sa covariance sont définies à partir de \eqref{eq:mesure} pour quantifier l'inadéquation entre le vecteur d'état prédit et la mesure, comme suit :
            \begin{subequations}
                \begin{align}
                    \label{eq:innovation}
                    \widehat{\mathbf{y}}_t & = \mathbf{z}_t - \mathbf{H}_t \widehat{\mathbf{x}}_{t \vert t-1},\\
                \label{eq:covariance_innovation}
                    \mathbf{S}_t & = \mathbf{H}_t \mathbf{P}_{t \vert t-1} \transpose{\mathbf{H}_t} + \mathbf{R}_t.
                \end{align}
            \end{subequations}
            Ensuite, un gain est appliqué à l'innovation pour corriger le vecteur d'état prédit et la covariance de l'erreur, selon les relations suivantes : 
            \begin{subequations}
                \begin{align}
                    \label{eq:correction}
                    \widehat{\mathbf{x}}_{t \vert t} & = \widehat{\mathbf{x}}_{t \vert t-1} + \mathbf{K}_t \widehat{\mathbf{y}}_t, \\
                    \label{eq:covariance_correction}
                    \mathbf{P}_{t \vert t}         & = (\mathbf{I} - \mathbf{K}_t \mathbf{H}_t) \mathbf{P}_{t \vert t-1}.
                \end{align}
            \end{subequations}
            Ce gain est calculé analytiquement à partir de la covariance de l'innovation afin de minimiser l'erreur quadratique moyenne du vecteur d'état corrigé, comme suit :      
            \begin{equation}
                \label{eq:gain_kalman}
                \mathbf{K}_t = \mathbf{P}_{t \vert t-1} \transpose{\mathbf{H}_t} {\mathbf{S}_t}^{-1}.
            \end{equation}
            L'expression compacte \eqref{eq:covariance_correction} de la covariance de l'erreur après correction découle de simplifications algébriques permises par la structure du gain de Kalman. De plus, un calcul direct montre que le gain de Kalman est spécifiquement conçu pour décorréler l'erreur de l'innovation, ce qui, en présence de bruits gaussiens, revient à les rendre indépendants.

        \subsection{Filtres de Kalman profonds}
        \label{sec:kalman_profonds}

            Les conditions d'optimalité du filtre de Kalman sont rarement remplies dans les applications réelles, ce qui peut fortement dégrader ses performances, qui dépendent également de la précision des modèles. Par exemple, en navigation, les bruits de mesure des systèmes de positionnement par satellite présentent souvent des corrélations temporelles, ne sont presque jamais gaussiens, et leurs covariances sont difficilement accessibles.
            
            Avec l'avènement des méthodes d'apprentissage profond, diverses architectures de réseaux de neurones récurrents guidés par un filtre de Kalman ont été développées pour pallier les limitations de ce dernier. Ces modèles reposent principalement sur une architecture commune, le KalmanNet \cite{KalmanNet}, qui remplace le calcul analytique du gain de Kalman par un apprentissage basé sur les données. Cependant, les premiers modèles \cite{SplitKalmanNet} ne permettent pas d'estimer la covariance de l'erreur, ce qui est pourtant essentiel dans de nombreuses applications pratiques, telles que la navigation.
            
            Le Cholesky KalmanNet (CKN) \cite{CholeskyKalmanNet}, récemment introduit, permet d'estimer à la fois les états et la covariance des erreurs. Le gain est appris sous la forme \eqref{eq:gain_kalman} à l'aide de deux unités récurrentes fermées : l'une pour estimer la covariance des erreurs à la prédiction, et l'autre pour estimer l'inverse de la covariance de l'innovation. Son entraînement est néanmoins complexe. Il repose sur une optimisation alternée de ces deux unités, avec une fonction coût qui pénalise indépendamment, et de manière arbitraire selon le choix d'un hyperparamètre, le gain (via l'erreur quadratique moyenne) et la covariance des erreurs après corrections (via l'écart moyen à la covariance empirique des erreurs).

    \section{Recursive KalmanNet (RKN)}
    \label{sec:recursive_kalmannet}

        Nous avons introduit le Recursive KalmanNet (RKN) \cite{RecursiveKalmanNet}, un réseau de neurones récurrents couplé à un filtre de Kalman, conçu pour estimer les variables d'état ainsi que la covariance des erreurs d'estimation d'un système dynamique stochastique à partir de mesures bruitées, sans nécessiter de connaissance préalable des covariances du bruit. Cet estimateur préserve la structure du filtre de Kalman tout en apprenant le gain et la covariance de l'erreur grâce à l'apprentissage profond.
        
        Contrairement au CKN, le RKN estime la covariance $\mathbf{P}_{t \vert t}$ de l'erreur sur la correction sans celle de la prédiction $\mathbf{P}_{t \vert t-1}$.  Cette méthode empêche de déterminer le gain selon \eqref{eq:gain_kalman} et d'appliquer \eqref{eq:covariance_correction}. La première étape consiste donc à utiliser la formule générale de Joseph \cite{Joseph} pour exprimer $\mathbf{P}_{t \vert t}$ :
        \begin{equation}
            \label{eq:joseph}
            \mathbf{P}_{t \vert t} = (\mathbf{I} - \mathbf{K}_t \mathbf{H}_t) \mathbf{P}_{t \vert t-1} \transpose{(\mathbf{I} - \mathbf{K}_t \mathbf{H}_t)} + \mathbf{K}_t \mathbf{R}_t \transpose{\mathbf{K}_t}.
        \end{equation}
        La dépendance à la covariance de l'erreur sur la prédiction est éliminée en injectant l'équation de propagation \eqref{eq:covariance_prediction}, ce qui permet de réécrire la covariance de l'erreur sur la correction comme la somme $\mathbf{P}_{t \vert t} = \mathbf{A}_t + \mathbf{B}_t$ de deux contributions :
        \begin{subequations}
            \begin{align}
                \label{eq:covariance_propag}
                \mathbf{A}_t & = (\mathbf{I} - \mathbf{K}_t \mathbf{H}_t) \mathbf{F}_t \mathbf{P}_{t - 1 \vert t-1} \transpose{\mathbf{F}_t} \transpose{(\mathbf{I} - \mathbf{K}_t \mathbf{H}_t)}, \\
                \label{eq:covariance_bruit}
                \mathbf{B}_t & = (\mathbf{I} - \mathbf{K}_t \mathbf{H}_t) \mathbf{Q}_t \transpose{(\mathbf{I} - \mathbf{K}_t \mathbf{H}_t)} + \mathbf{K}_t \mathbf{R}_t \transpose{\mathbf{K}_t}.
            \end{align}
        \end{subequations}
        Le premier terme dépend de la covariance apprise à l'instant précédent et du gain, tandis que le second terme dépend du gain et des covariances des bruits, qui sont inconnues du RKN.

        L'architecture générale du RKN est illustrée à la Figure \ref{fig:RKN_architecture}. Elle se compose de deux unités récurrentes fermées (GRU), entourées de réseaux de neurones entièrement connectés qui prennent en entrée l'innovation, la correction précédente, la matrice de mesure et la différence entre les deux dernières mesures. Les entrées ne sont pas normalisées par lots et sont mises au carré en pratique pour accélérer l'apprentissage, car les sorties à estimer sont quadratiques par rapport aux entrées. La première unité estime le gain $\widehat{\mathbf{K}}_t$, tandis que la seconde calcule le facteur de Cholesky $\widehat{\mathbf{C}}_t$ de \eqref{eq:covariance_bruit}.

        L'apprentissage du RKN s'effectue sur des ensembles constitués de séries temporelles de vecteurs d'état et de mesure. Il repose sur une descente de gradient, appliquée à la moyenne de la log-vraisemblance négative gaussienne de l'erreur sur la correction, calculée à la fois sur les lots et dans le temps, ce qui permet de pénaliser simultanément le gain et la covariance des erreurs. Une normalisation $\ell^2$ est également appliquée aux paramètres des réseaux de neurones, voir \cite[Équation (9)]{RecursiveKalmanNet}\footnote{Code du RKN sur GitHub : \href{https://github.com/ixblue/RecursiveKalmanNet}{github.com/ixblue/RecursiveKalmanNet}.}. 

        Le calcul du gradient de la fonction de coût du RKN par rapport aux sorties, \cite[Équations (10a) et (10b)]{RecursiveKalmanNet}, montre que les covariances estimées sont statistiquement représentatives des erreurs sur le vecteur d'état. Le gain appris décorrèle erreur et innovation, ce qui est un prérequis pour que toute l'information de la mesure soit incorporée dans la correction.

    \section{Degré de généralisation du RKN}
    \label{sec:generalisation}

        En navigation inertielle, dès que le cap initial est estimé, ce qui est généralement effectué à l'issue d'une phase d'alignement en gyrocompas, les équations de navigation sont linéarisables. Nous évaluons ainsi les capacités de généralisation de notre modèle sur un système cinématique unidimensionnel à vitesse constante, où les états sont la position et la vitesse, et la mesure correspond à la position. L'espace d'état est donné par :
        \begin{alignat}{2}
            \label{eq:exemple_propagation}
            \mathbf{x}_t & = \begin{pmatrix}
                                1 & \dt \\
                                0 & 1
                             \end{pmatrix} \mathbf{x}_{t-1} + \mathbf{v}_t, 
                                & \quad & \mathbf{x}_t \in \mathbb{R}^2, \\
            \label{eq:exemple_mesure}
            z_t          & = \begin{pmatrix}
                                1 & 0
                             \end{pmatrix} \mathbf{x}_t + w_t, 
                                & \quad & z_t \in \mathbb{R}.
        \end{alignat}
        Nous travaillons avec un bruit de propagation issu d'un bruit blanc gaussien en accélération, intégré en une marche aléatoire de vitesse dont les incréments sont normalement distribués, de moyenne nulle et d'écart-type $\sigma_v$, à savoir
        $\mathbf{Q}_t = \begin{pmatrix}
                            0 & 0 \\
                            0 & {\sigma_v}^2
                        \end{pmatrix}$. 
        Notre bruit de mesure est un bruit blanc gaussien de moyenne nulle et de covariance $\mathbf{R}_t = {\sigma_t}^2$. Dans les simulations, nous fixons $\dt = 1$ et $\sigma_v = 0.01$, sans préciser d'unité.

        \subsection{Scénarios d'entraînement}
        \label{sec:entrainement}

            En raison de la grande diversité des bruits de mesure rencontrés en navigation inertielle, nous choisissons d'évaluer la capacité de généralisation du RKN dans des scénarios où le modèle de génération des bruits de mesure diffère entre les données d'entraînement et celles de test.

            Pour entraîner le RKN, nous construisons trois ensembles de données, chacun comprenant $1000$ paires de séries temporelles de longueur $150$ pour l'entraînement et $100$ paires de même longueur pour la validation, chaque paire étant constituée des états et des mesures correspondantes.

            Les conditions initiales des vecteurs d'état suivent une loi normale, de moyenne et covariance respectivement égales à :
            \[
                \widehat{\mathbf{x}}_{0 \vert 0} = 
                    \begin{pmatrix} 
                        0 \\ 
                        1
                    \end{pmatrix},
                \quad
                \mathbf{P}_{0 \vert 0} = 
                    \begin{pmatrix}
                        1 & 0 \\
                        0 & 0.01
                    \end{pmatrix}.
            \]                     
            Leur évolution temporelle est ensuite déterminée par \eqref{eq:exemple_propagation}, avec des tirages de bruit blanc gaussien de moyenne nulle et de covariance $\mathbf{Q}_t$. Les réalisations de ces deux tirages aléatoires diffèrent entre les trois ensembles de données.
            
            Les mesures sont générées selon \eqref{eq:exemple_mesure} avec des tirages de bruit blanc gaussien de moyenne nulle et de covariance $\mathbf{R}_t = {\sigma_t}^2$, dont la structure varie selon les scénarios d'entraînement.

            \subsubsection*{Scénario 1 -- Changement brutal de régime}
            
                Un premier modèle, nommé $\text{RKN}_{\text{Réf}}$, est entraîné sur des bruits de mesure dont l'écart type présente une transition abrupte identique à un même instant. Concrètement, $\sigma_t$ est donné par :
                \[
                    \sigma_t = 
                        \begin{cases}
                            0.35, & t \in [0, 75[,\\
                            1.75, & t \in [75, 150].
                        \end{cases}
                \]
                Le $\text{RKN}_{\text{Réf}}$ est ainsi nommé, car il est entraîné sur des données issues de la distribution de test présentée en Section \ref{sec:methodologie}.

            \subsubsection*{Scénario 2 -- Deux régimes constants proches}
            
                Un deuxième modèle, nommé $\text{RKN}_{\text{E}1}$, est entraîné sur des bruits de mesure d'écart type constant, répartis équitablement entre $\sigma_t = 1.5$ et $\sigma_t = 0.6$.

            \subsubsection*{Scénario 3 -- Deux régimes constants éloignés}

                Un troisième modèle, nommé $\text{RKN}_{\text{E}2}$, est entraîné sur des bruits de mesure d'écart type constant, répartis équitablement entre $\sigma_t \approx 1.91$ et $\sigma_t \approx 0.19$.\\
                
            Les écarts types en régime constant sont choisis pour être de rapports respectifs $5$, $2.5$ et $10$ en Scénario 1, 2 et 3.

            \begin{figure*}[htbp]
                \centering
                \subfloat[Comparaison du RKN aux filtres de Kalman.]{\includegraphics[width=0.33\linewidth]{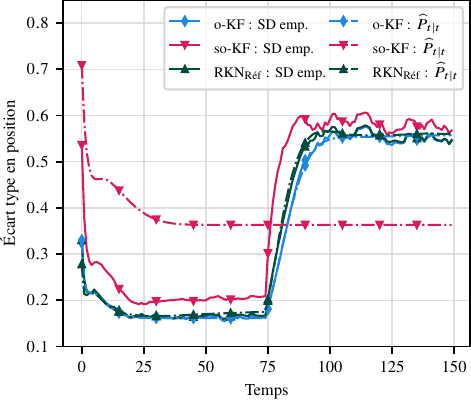}
                \label{fig:sd_estimateurs}} 
                \subfloat[Comparaison des entraînements du RKN.]{\includegraphics[width=0.33\linewidth]{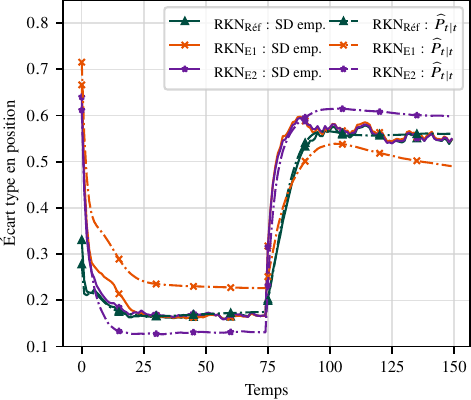}
                \label{fig:sd_entrainements}}
                \subfloat[Comparaison des gains moyens en position.]{\includegraphics[width=0.329\linewidth]{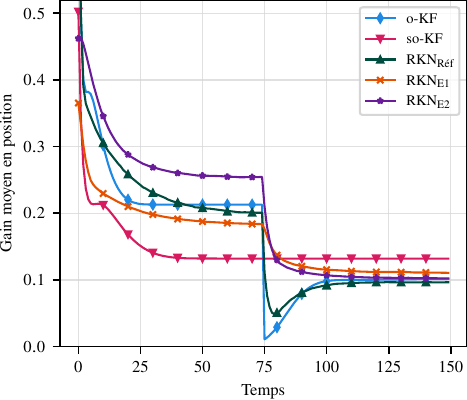}
                \label{fig:gain}}
                \caption{Comparaison en position des estimateurs : moyennes quadratiques sur les tests des écarts types estimés (lignes continues) et empiriques (lignes tiret-point) dans les deux premières figures, et gains moyens sur les tests dans la dernière.}
                \label{fig:comparaison_estimateurs}
            \end{figure*}

        \subsection{Critères de généralisation}
        \label{sec:methodologie}

            Les performances de généralisation du RKN sont évaluées sur un ensemble de test composé de $1000$ paires de séries temporelles de longueur $150$, générées de la même manière qu'en entraînement pour $\text{RKN}_{\text{Réf}}$ (Scénario 1, Section \ref{sec:entrainement}).

            Nous comparons d'abord $\text{RKN}_{\text{Réf}}$ au filtre de Kalman sous deux versions : une optimale (o-KF), où la covariance des bruits de mesure est modélisée à partir de celle utilisée pour générer les données de test, et une sous-optimale (so-KF), où cette covariance est, par convention, fixée à $1$. Nous évaluons ensuite les apprentissages en comparant les différents $\text{RKN}$, en prenant $\text{RKN}_{\text{Réf}}$ comme référence, puisqu'il est entraîné sur des données similaires à celles de test.

            Nous mesurons respectivement la précision des estimateurs et la représentativité des covariances en termes des normes quadratiques moyennes des erreurs, normalisées ou non, soit :
            \begin{align*}
                \text{EQM}(t)  & = \frac{1}{N}\sum_{k = 1}^N \transpose{\left(\mathbf{x}_t^{(k)} - \widehat{\mathbf{x}}_t^{(k)}\right)}\left(\mathbf{x}_t^{(k)} - \widehat{\mathbf{x}}_t^{(k)}\right), \\
                \text{EQM}_{\text{n}}(t) & = \frac{1}{N}\sum_{k = 1}^N\transpose{\left(\mathbf{x}_t^{(k)} - \widehat{\mathbf{x}}_t^{(k)}\right)}{\mathbf{P}_{t \vert t}^{(k)}}^{-1}\left(\mathbf{x}_t^{(k)} - \widehat{\mathbf{x}}_t^{(k)}\right),
            \end{align*}
            où $N = 1000$ est la taille de l'ensemble de test, et $\mathbf{x}_t^{(k)}$ et $\widehat{\mathbf{x}}_t^{(k)}$ sont respectivement l'état réel et estimé sur le $k$-ème lot.

            Si les erreurs $\mathbf{x}_t^{(k)} - \widehat{\mathbf{x}}_t^{(k)}$ sont gaussiennes et indépendantes, comme c'est le cas pour o-KF, alors le théorème central limite appliqué à une loi du khi-deux dont le nombre de paramètres est égal au nombre d'états $m$, montre que $\text{EQM}_{\text{n}}(t)$ suit une loi normale de moyenne $m = 2$ et de variance $2m/N = 0.004$, fournissant ainsi un attendu théorique.
            
        \subsection{Résultats des entraînements}
        \label{sec:resultats}

            La Table \ref{tab:comparaison_performance} compare les EQM et $\text{EQM}_{\text{n}}$ des estimateurs à l'asymptotique du premier régime de bruit de mesure et lors du changement de régime. Les Figures \ref{fig:sd_estimateurs} et \ref{fig:sd_entrainements} présentent les écarts-types de position estimé et empirique typiques : plus la ligne continue est basse, plus l'estimateur est précis ; plus elle se rapproche de la ligne tiret-point, plus l'écart-type estimé reflète fidèlement la statistique des erreurs empiriques.
            
            \begin{table}[h]
                \centering
                \begin{tabular}{|c||c|c|c|c|c|}
                    \hline 
                    & \multicolumn{2}{|c|}{$t = 70$} & \multicolumn{2}{|c|}{$t = 80$} \\
                    \hline 
                                                        & $\text{EQM}(t)$ & $\text{EQM}_{\text{n}}(t)$ & $\text{EQM}(t)$ & $\text{EQM}_{\text{n}}(t)$ \\
                    \hline 
                    \textbf{o-KF}                       & -14             & 2.1                        & -9.7            & 1.9 \\
                    \hline
                    \textbf{so-KF}                      & -12             & 1.1                        & -4.9            & 2.5 \\
                    \hline
                    \textbf{$\text{RKN}_{\text{Réf}}$}  & -14             & 1.9                        & -8.1            & 2.1 \\
                    \hline
                    \textbf{$\text{RKN}_{\text{E1}}$}   & -14             & 1.4                        & -4.2            & 3.3 \\
                    \hline
                    \textbf{$\text{RKN}_{\text{E2}}$}   & -14             & 3.2                        & -4.3            & 2.8 \\
                    \hline
                \end{tabular}
                \caption{Comparaison des estimateurs en termes de EQM, exprimé en dB, et $\text{EQM}_{\text{n}}$, à l'asymptotique du premier régime de bruit de mesure et pendant le changement de régime.}
                \label{tab:comparaison_performance}
            \end{table}
                
            Le $\text{RKN}_{\text{Réf}}$ est systématiquement plus précis que le so-KF, et sa covariance est représentative des erreurs, contrairement à celle du so-KF, qui ne s'adapte pas aux changements de régime du bruit de mesure. Les performances du $\text{RKN}_{\text{Réf}}$ se rapprochent même de celles de l'o-KF, bien qu'il ignore l'écart type du bruit de mesure. Les $\text{RKN}_{\text{E1}}$ et $\text{RKN}_{\text{E2}}$ conservent une précision similaire à celle du $\text{RKN}_{\text{Réf}}$, mais n'apprennent que des covariances similaires à celles vues à l'entraînement.

            Pour mieux comprendre ces résultats et la dynamique des estimateurs, la Figure \ref{fig:gain} illustre l'évolution temporelle typique de leurs gains en position. Contrairement au so-KF, dont le gain ne réagit pas au changement d'écart type du bruit de mesure, tous les RKN réussissent à détecter ces variations, même sans y avoir été exposés en entraînement, et estiment des gains asymptotiques proches de ceux de l'o-KF. Cette capacité de détection et d’adaptation reste valide même en présence de transitions multiples de l'écart type du bruit de mesure. Des observations similaires sont faites en vitesse.
            
    \section{Conclusion}
    \label{sec:conclusion}

        Le Recursive KalmanNet généralise bien dans des scénarios où le bruit de mesure diffère entre l'entraînement et le test, tout en conservant une précision proche de celle du filtre de Kalman optimal. Les gains appris montrent sa capacité à modéliser des changements de régime, même en l'absence de ces derniers lors de l'entraînement. Cependant, les covariances apprises ne sont pas toujours représentatives des erreurs empiriques, ce qui suggère d'examiner la sortie du modèle qui estime la composante liée au bruit de la covariance.

    \renewcommand{\refname}{\footnotesize\bfseries Références}
    \footnotesize
    \bibliography{biblio}

\begin{thebibliography}{1}
\expandafter\ifx\csname fonteauteurs\endcsname\relax
\def\fonteauteurs{\scshape}\fi

\bibitem{Generalization}
Olivier \bgroup\fonteauteurs\bgroup Bousquet\egroup\egroup{} et Andr{\'e}
  \bgroup\fonteauteurs\bgroup Elisseeff\egroup\egroup{} :
\newblock {Stability and generalization}.
\newblock {\em J. Mach. Learn. Res}, 2\string:\penalty500\relax 499--526, 2002.

\bibitem{Joseph}
Richard~S. \bgroup\fonteauteurs\bgroup Bucy\egroup\egroup{} et Peter~D.
  \bgroup\fonteauteurs\bgroup Joseph\egroup\egroup{} :
\newblock {\em {Filtering for Stochastic Processes with Applications to
  Guidance}}, volume 326.
\newblock Am. Math. Soc., 2005.

\bibitem{SplitKalmanNet}
Geon \bgroup\fonteauteurs\bgroup Choi\egroup\egroup{}, Jeonghun
  \bgroup\fonteauteurs\bgroup Park\egroup\egroup{}, Nir
  \bgroup\fonteauteurs\bgroup Shlezinger\egroup\egroup{}, Yonina~C.
  \bgroup\fonteauteurs\bgroup Eldar\egroup\egroup{} et \hbox{Namyoon}
  \bgroup\fonteauteurs\bgroup Lee\egroup\egroup{} :
\newblock {Split-KalmanNet: A Robust Model-Based Deep \hbox{Learning} Approach
  for State Estimation}.
\newblock {\em IEEE Trans. Veh. Technol.}, 72\string:\penalty500\relax
  12326--12331, 2023.

\bibitem{Kalman}
Rudolph~E. \bgroup\fonteauteurs\bgroup Kalman\egroup\egroup{} :
\newblock {A New Approach to Linear Filtering and \hbox{Prediction} Problems}.
\newblock {\em J. Basic Eng.}, 82\string:\penalty500\relax 35--45, 1960.

\bibitem{CholeskyKalmanNet}
Minhyeok \bgroup\fonteauteurs\bgroup Ko\egroup\egroup{} et Abdollah
  \bgroup\fonteauteurs\bgroup Shafieezadeh\egroup\egroup{} :
\newblock {Cholesky-KalmanNet: Model-Based Deep Learning with Positive Definite
  Error Covariance Structure}.
\newblock {\em IEEE Signal Process. Lett.}, 32\string:\penalty500\relax
  326--330, 2025.

\bibitem{RecursiveKalmanNet}
Hassan \bgroup\fonteauteurs\bgroup Mortada\egroup\egroup{}, Cyril
  \bgroup\fonteauteurs\bgroup Falcon\egroup\egroup{}, Yanis
  \bgroup\fonteauteurs\bgroup Kahil\egroup\egroup{}, Mathéo
  \bgroup\fonteauteurs\bgroup Clavaud\egroup\egroup{} et Jean-Philippe
  \bgroup\fonteauteurs\bgroup Michel\egroup\egroup{} :
\newblock {Recursive KalmanNet: Deep Learning-Augmented Kalman Filtering for
  State Estimation with Consistent \hbox{Uncertainty} Quantification}.
\newblock \emph{In} {\em 2025 33rd Eur. Signal Process. Conf. (EUSIPCO)}. IEEE,
  2025.

\bibitem{KalmanNet}
Guy \bgroup\fonteauteurs\bgroup Revach\egroup\egroup{}, Nir
  \bgroup\fonteauteurs\bgroup Shlezinger\egroup\egroup{}, Xiaoyong
  \bgroup\fonteauteurs\bgroup Ni\egroup\egroup{}, Adria~Lopez
  \bgroup\fonteauteurs\bgroup Escoriza\egroup\egroup{}, Ruud J.~G.
  \bgroup\fonteauteurs\bgroup Van~Sloun\egroup\egroup{} et Yonina~C.
  \bgroup\fonteauteurs\bgroup Eldar\egroup\egroup{} :
\newblock {KalmanNet: Neural Network-Aided Kalman Filtering for Partially Known
  Dynamics}.
\newblock {\em IEEE Trans. Signal Process.}, 70\string:\penalty500\relax
  1532--1547, 2022.

\bibitem{Schmidt}
Stanley~F. \bgroup\fonteauteurs\bgroup Schmidt\egroup\egroup{} :
\newblock {Application of State-Space Methods to \hbox{Navigation} Problems}.
\newblock \emph{In} Cornelius~T. \bgroup\fonteauteurs\bgroup
  Leondes\egroup\egroup{}, \'editeur :  {\em Adv. Control Syst.}, volume~3,
  pages 293--340. Elsevier, 1966.

\end{thebibliography}
    
\end{document}